\documentclass[aps,prl,twocolumn,showpacs]{revtex4}
\usepackage{graphicx}
\begin{document}
%
%
\title{Screened Vortex Lattice Model with Disorder}
\author{C. Giardin\`a$^{1,2}$,  N. V. Priezjev$^{2}$,
J. M. Kosterlitz$^{2}$}
\affiliation{$^{1}$ Dipartimento di Matematica, Universit\`a
di Bologna, 40127 Bologna, Italy}
\affiliation{$^{2}$Department of Physics, Brown University,
Providence, RI, 02912, USA}
\begin{abstract}
The three dimensional $XY$ model with quenched random disorder and finite screening
is studied. We argue that the system scales to model with $\lambda\simeq 0\simeq T$
and the resulting effective model is studied numerically by defect energy scaling.
In zero external field we find that there exists a true superconducting
phase with a stiffness exponent $\theta\simeq +1.0$ for weak disorder.
For low magnetic field and weak disorder, there is also a superconducting phase with
$\theta\simeq +1.0$ which we conjecture is a Bragg glass.
For larger disorder or applied field, there is a non superconducting
phase with $\theta\simeq -1.0$. We estimate the critical external field
whose value is consistent with experiment.
\end{abstract}
\maketitle
\begin{figure}
\includegraphics[width=4.0cm, height=4.0cm]{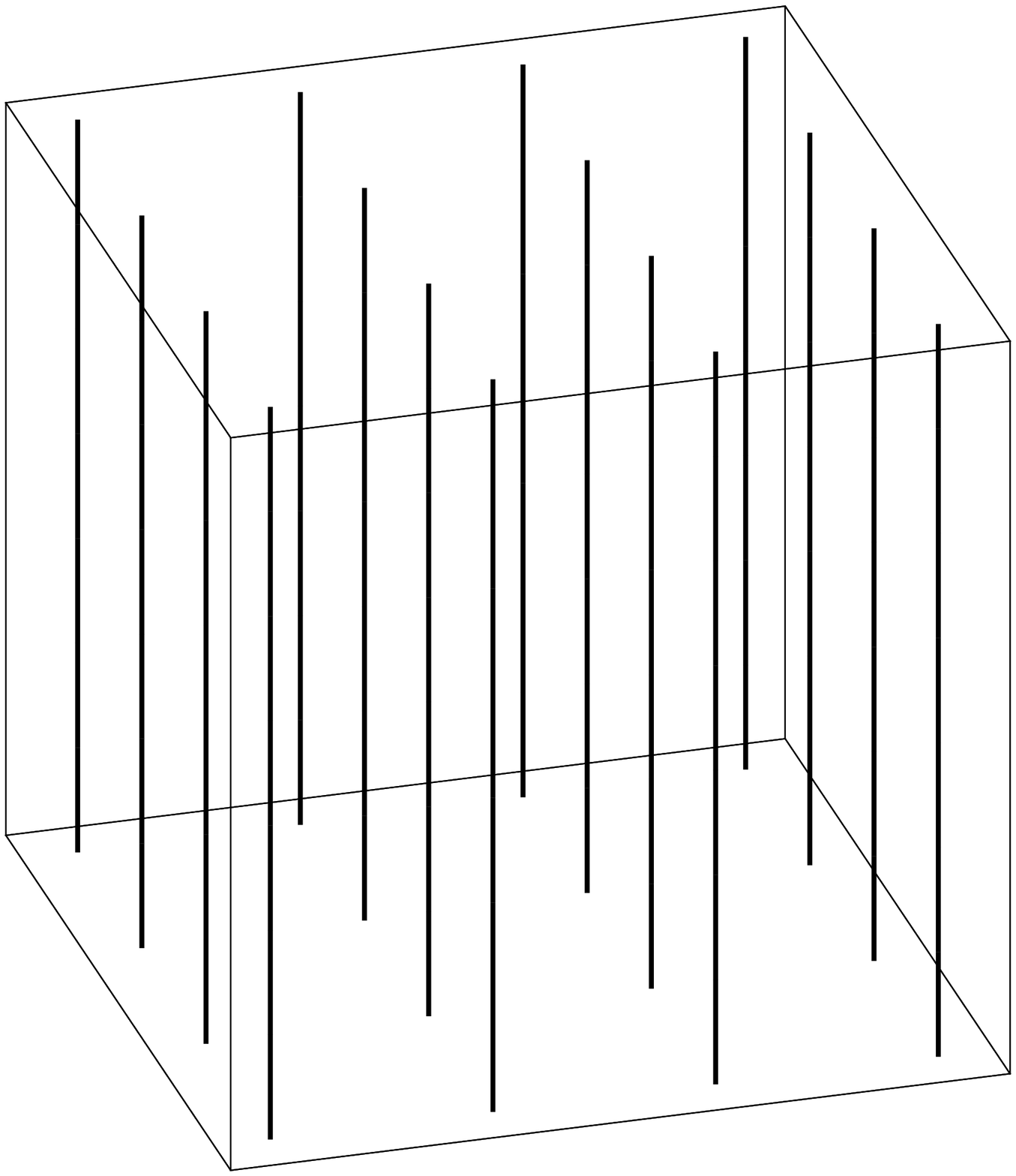}
\includegraphics[width=4.0cm, height=4.0cm]{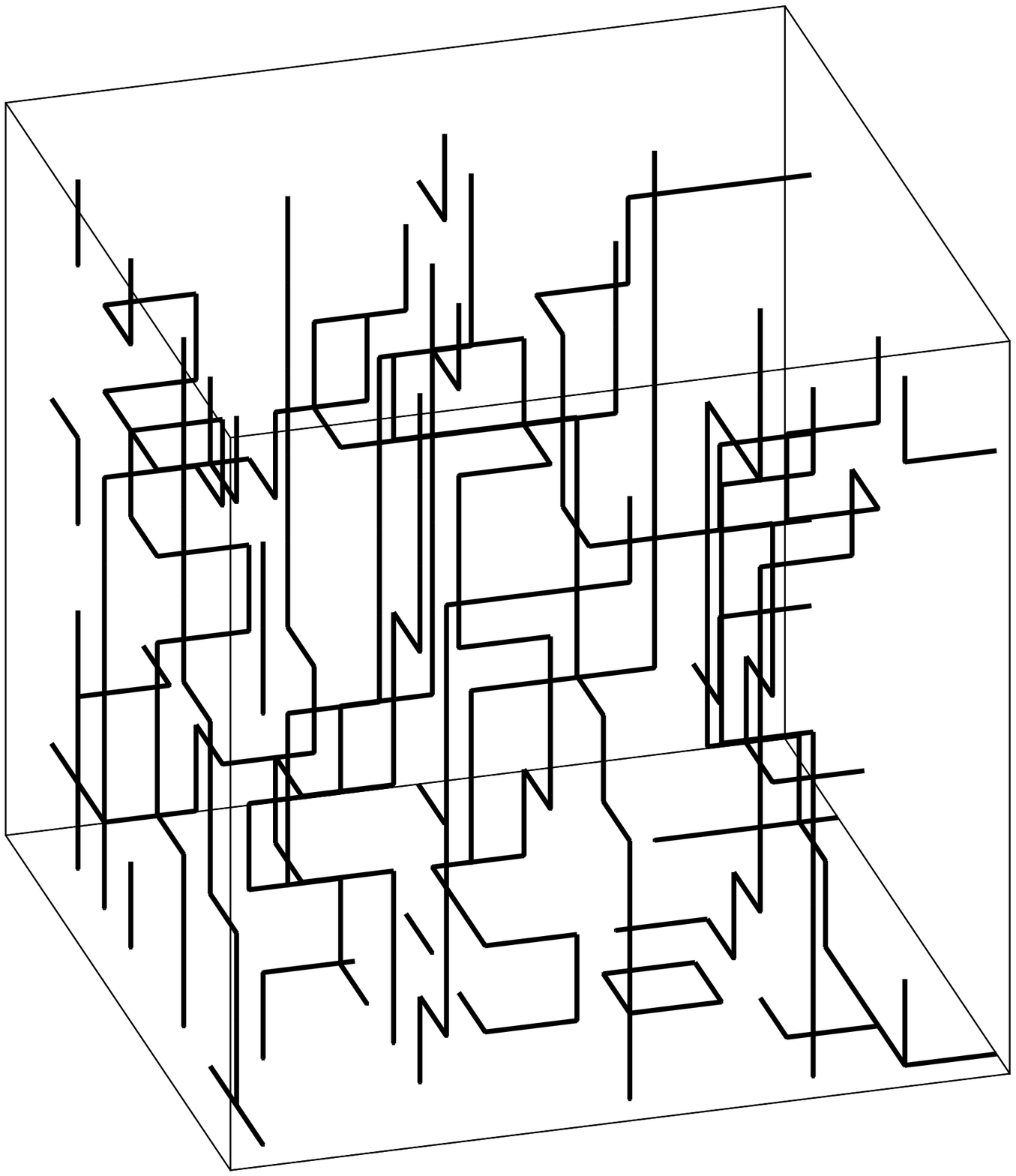}
\caption{Ground states for $h=0.25$ in the ordered phase, $\alpha=0.10$
(left) and in the disordered phase, $\alpha=0.48$ (right).}
\label{ground_state}
\end{figure}
\begin{figure}
\includegraphics[width=6.0cm,  angle=-90]{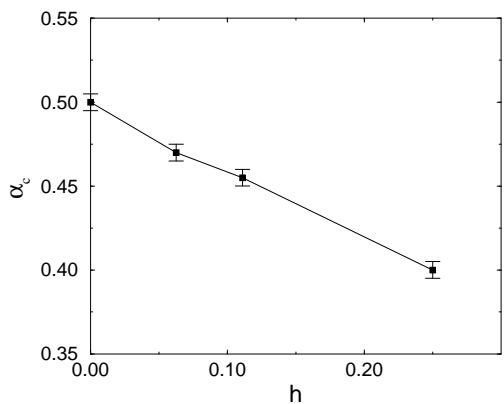}
\caption{Critical disorder strength $\alpha_{c}(h)$ for $0\le h\le 0.25$.}
\label{alpha.vs.field}
\end{figure}

The phase diagram of type II superconductors in an external field has been the subject of
intense theoretical and experimental investigation \cite{Blatter:94;Natterman:00}.
After the discovery of high-$T_{c}$ materials, the role of both thermal and disorder induced
fluctuations has been reconsidered, revealing many new interesting phenomena. In clean systems,
it was realized \cite{Nelson:88;Houghton:89} that, with increasing temperature, the Abrikosov
lattice melts into a vortex liquid via a thermally induced first-order transition. Experiments
performed on thermodynamic quantities such as magnetization \cite{Zeldov:95} and specific heat
\cite{Schilling:96} confirmed the first order nature of the melting transition in YBCO and BSSCO
materials. More recent studies of the clean unscreened system in an external field
\cite{nguyen:99;hove:99} have shown that the low temperature vortex lattice melts at $T=T_{M}$ to
a liquid of rigid flux lines and at $T_{L}\geq T_{M}$ the lines become entangled and vortex loops
proliferate. Very recently, there have been studies \cite{nono:01,olsson:01,vestergren:02} of
a random $3D$ $XY$ model wthout screening. These claim to study the glass transition in
a superconductor but do not have the vital screening and obtain results which are inconsistent
with each other so their relevance to real systems is unclear.
\newline\indent
In this Letter, we study the stability of superconductivity in the presence of point
disorder and an applied magnetic field by computing numerically the stiffness exponent
$\theta$. We argue that the effective screening length $\lambda$ and temperature $T$
scale to zero at very long length scales so that a model with $\lambda\simeq 0\simeq T$,
which is amenable to simulation, is a physically reasonable model of a disordered
superconductor at low $T$. Using this effective model, we can identify the field and disorder
driven transition \cite{Cubitt:93;Klein:01} from a superconducting to non-superconducting phase
as the field or disorder is increased and conjecture that the former is a Bragg glass
\cite{Giamarchi:94;Giamarchi:95}. The nature of the large disorder phase is still
controversial: a viscous non-superconducting pinned vortex liquid \cite{Giamarchi:97}
or a superconducting vortex glass \cite{Fisher:89;Fisher:91}.
\newline\indent
The ingredients necessary to describe a typical high-$T_{c}$ superconductor in a field are
(i) pinning of flux or vortex lines and loops by random point impurities and (ii) weak
screening of the interactions between vortices. We argue that a model containing these
essential ingredients is a three dimensional $XY$ model on a simple cubic lattice with quenched
random phase shifts. In the vortex representation, the Hamiltonian is
\cite{{Villain:75},{Jose:77},{Vallat:94}}
\begin{equation}
\label{vortex}
H = \frac{1}{2}\sum_{i,j}G(i,j)(\mbox{\boldmath $J$}_{i} - \mbox{\boldmath $b$}_{i})
\mbox{\boldmath $\cdot$}(\mbox{\boldmath $J$}_{j} - \mbox{\boldmath $b$}_{j})
\end{equation}
We may ignore boundary terms which, at least in the best twist approach, vanish by a proper
choice of global twists \cite{Kosterlitz:98}. The dynamical variables are the integer valued
vorticities $\mbox{\boldmath $J$}_{i}$ on the links of the dual lattice and subject to the
local constraint $(\nabla \cdot \mbox{\boldmath $J$})_{i} = 0$ at every site $i$.
The $\mbox{\boldmath $b$}_i$ are quenched random fluxes on the dual lattice which are obtained
from the circulation of the quenched vector potential $\mbox{\boldmath $A$}$ and by adding a
uniform external field $h$ in the $\mbox{\boldmath $\hat{z}$}$ direction
\begin{equation}
\label{b}
\mbox{\boldmath $b$}_{i} = \frac{1}{2\pi} [\nabla \times
\mbox{\boldmath $A$}]_{i} +  h \mbox{\boldmath $\hat{z}$}
\end{equation}
Here $h\equiv Ba_{0}^{2}/\Phi_{0}$ is the mean flux per elementary plaquette normal to the applied
field $B\hat{\bf z}$, $a_{0}$ is the underlying lattice spacing and $\Phi_{0}=2\times 10^{-7}$
Gauss$\cdot$cm$^{2}$ is the flux quantum. The vector potential $A_{\mu i}$ with $\mu=x,y,z$ is
independently uniformly distributed $A_{\mu i}\in [0,2\pi\alpha)$ with $0\le\alpha\le 1$ and is
defined on the bonds of the original lattice. The {\em disorder strength $\alpha$} interpolates
between two well known limits, the clean system, $\alpha = 0$, and the maximally disordered,
$\alpha =1$, gauge glass. By construction, the fields $\mbox{\boldmath $b$}_{i}$  satisfy the
divergenceless condition $(\nabla \cdot \mbox{\boldmath $b$})_{i} = 0$ on every site. $G(i,j)$ is
the screened lattice Green's function with dimensionless screening length $\lambda$ in units of $a_{0}$
\begin{equation}\label{eq:gf}
G(i,j) = \frac{(2\pi)^2}{L^3}\sum_{\bf k}
\frac{\exp [i\mbox{\boldmath $k$} \cdot
(\mbox{\boldmath $r$}_i - \mbox{\boldmath $r$}_j)]}
{2\sum_{\mu}(1-\cos k_{\mu}) + \lambda^{-2}}
\end{equation}
where $\mbox{\boldmath $r$}_i = (x_i,y_i,z_i)$ is the $i$-th site on the dual lattice and
$k_{\mu} = 2\pi n_{\mu}/L$, with $\mu = x,y,z$ and $n_{\mu}=(1,\ldots,L)$.
The vortex system defined by Eqs. (\ref{vortex}), (\ref{b}) and (\ref{eq:gf}) is a vortex
lattice of spacing $h^{-1/2}$ for $\alpha =0$ in the absence of disorder. Turning on disorder
distorts the lattice and pins it at a favorable position in the underlying substrate lattice,
as is clear from Eq. (\ref{vortex}), but does not destroy the underlying periodicity of the
vortex structure \cite{Giamarchi:94;Giamarchi:95}. Real cuprate superconductors are very
anisotropic with $\Gamma = \lambda_{z}/\lambda_{x,y} \approx 5$ for YBCO and $\Gamma \approx 100$
for BSSCO \cite{Blatter:94;Natterman:00} where we have chosen the $\mbox{\boldmath $\hat{z}$}$
direction as the $c$ axis and $\mbox{\boldmath $\hat{x},\hat{y}$}$ directions as the $a,b$ axes.
We note that when $\lambda_{\mu}\ll 1$, the vortex-vortex interactions become isotropic
\cite{nsh96} and, in this limit, our isotropic model is realistic.
\newline\indent
The final and most important part of our argument is to justify the
validity of the strong screening limit for an extreme type II superconductor where
the bare screening length $\infty >\lambda\gg 1$ is large but finite. >From Eq. (\ref{eq:gf}), it
is clear that $\lambda$ scales like a length so that scaling $a_{0}\rightarrow e^{l}a_{0}$
induces the scaling $\lambda\rightarrow e^{-l}\lambda$. Thus, at long length scales which are
relevant to the weakly disordered system of interest, the effective screening length
$\lambda\ll 1$. In this strong screening limit, $G(i,j)=(2\pi\lambda )^{2}\delta_{ij}+
{\cal O}(\lambda^{4})$ which yields
\begin{equation}
\label{Hamil_simpl}
\frac{H}{(2\pi\lambda)^{2}}  = \frac{1}{2}\sum_{i}(\mbox{\boldmath $J$}_{i} - \mbox{\boldmath $b$}_{i})^{2}
+ {\cal O}(\lambda^{2})
\end{equation}
This local form of Eq. (\ref{Hamil_simpl}) on a cubic lattice can be regarded as a model for the
long length scale properties of a disordered superconductor and can be studied numerically by
very efficient combinatorial optimization algorithms \cite{Ahuja:93,Rieger:98}
on large systems. Any nonlocal terms such as vortex - vortex interactions
or boundary terms \cite{Bokil:95,Kosterlitz:98} in Eq. (\ref{vortex})
render such algorithms useless. Of course, screening is weak, $\lambda\gg 1$,
in a real type II superconductor and one may question the relevance of a system
described by Eq. (\ref{Hamil_simpl}). Since our interest is in the superconductivity
and in the vortex lattice structure as a secondary effect, we argue that
Eq. (\ref{Hamil_simpl}) is an adequate description of a superconductor in an external
field with any finite screening. Numerical simulations \cite{Kosterlitz:98,Bokil:95,Wenegel:96} of
Eq. (\ref{vortex}) indicate that the exponent $\theta$ has the same value for
{\it any} $\lambda<\infty$. We find this slightly surprising as decreasing $\lambda$ transforms
a type II into a type I superconductor and the independence of $\theta$ on $\lambda$ seems to
imply that these do not differ in any essential way at small $T$. Additional justification for the
relevance of the model of Eq. (\ref{Hamil_simpl}) at $T=0$ is that $T$ is an irrelevant
variable \cite{Giamarchi:94;Giamarchi:95} scaling to zero in the superconducting region.
This gives some justification for the computationally accessible model of Eq. (\ref{Hamil_simpl})
and for our scaling arguments.
\newline\indent
To investigate whether a transition occurs, we use defect energy scaling.
In this approach, one computes the energy $\Delta E (L)$
of a defect in a system of linear size $L$ and fits to the scaling {\it ansatz}
\begin{equation}
\langle\Delta E (L)\rangle \sim L^{\theta}
\end{equation}
where $\theta$ is the stiffness exponent and $\langle\cdots\rangle$ denotes an
average over disorder. The sign of $\theta$ distinguishes two regimes:
if $\theta >0$, inserting a defect costs an infinite energy in the thermodynamic limit
and the system will be ordered at sufficiently small finite $T$. Conversely, if $\theta <0$,
large domains cost little energy and, at any $T>0$, superconductivity will be destroyed.
To calculate the defect energy we employ the method proposed by Kisker and Rieger \cite{Kisker:98},
who restated the problem of finding the ground state for Hamiltonian (\ref{Hamil_simpl})
in terms of a minimum-cost-flow problem \cite{Rieger:98},where the cost functions are precisely given
by $c_i(\mbox{\boldmath $J$}_i) = (\mbox{\boldmath $J$}_{i} - \mbox{\boldmath $b$}_{i})^{2}/2$.
This method makes use of the successive shortest path algorithm (SSPA) \cite{Ahuja:93} to
find the ground state configuration $\{ \mbox{\boldmath $J$}^0\}$ for each realization of disorder.
The global flux $\mbox{\boldmath $f$}$ associated with this configuration is given by
\begin{equation}
\label{flux}
\mbox{\boldmath $f$} = \frac{1}{L} \sum_i \mbox{\boldmath $J$}_i^{0}
\end{equation}
\newline\indent
The elementary low energy excitation configuration $\{ \mbox{\boldmath $J$}^1\}$ is obtained
by gradually decreasing all costs in, say, the $z$ direction until the global flux $f_{z}$
jumps by one, $f_{z} \rightarrow f_{z} + 1$. The lowest energy excitation will be a global
vortex loop encircling the $3D$ torus in the $z$ direction. The defect energy is then obtained by
$\Delta E = E(\mbox{\{\boldmath $J$}^{1}\}) -E(\mbox{\{\boldmath $J$}^{0}\})$.
A more conventional way of determining the defect energy is to calculate the
energy difference between periodic and antiperiodic boundary conditions, which amounts to adding
a global twist of $\pi$ along one spatial direction \cite{Kosterlitz:98}. In our case we can ignore
the boundary terms because the global loop corresponds to a twist of $2\pi$, which has no effect as
the original Hamiltonian is invariant under a discrete gauge transformation modulo $2\pi$.
Remarkably large system sizes can be treated by applying the SSPA, while conventional
methods such as repeated quenching or simulated annealing are much less efficient.
In this work we study $L\times L\times L$ systems with $L\leq 40$
for different values of $\alpha$ in the range $0\leq\alpha\leq 1$
and magnetic field in the range $0\leq h \leq 0.25$. The number of realizations of the random
bonds varies from $500$ for $40^3$ systems up to $10^{4}$ for the smallest ones.
There is no upper limit on $L$.
\begin{figure}
\includegraphics[width=7.0cm, height=7.0cm, angle=-90]{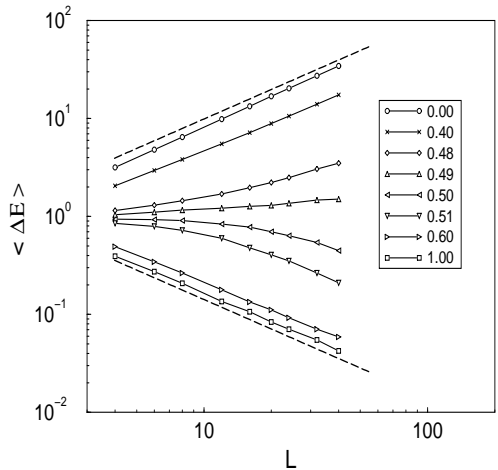}
\caption{Size $L$ dependence of domain wall energy $\langle\Delta E\rangle$
in zero external field (log-log plot). The legend shows the magnitude
of the disorder strength. Solid lines are guides for the eyes, dashed
lines with slopes $\pm 1$ are drawn for reference. }
\label{zerofield}
\end{figure}
\newline\indent
The zero field case has been studied in detail \cite{Pfeiffer:01}
and we repeated the simulation to check our algorithm. The results are
summarized in Fig. (\ref{zerofield}) for several values of
$\alpha$ and they are essentially identical to those of Ref. \cite{Pfeiffer:01}. For weak
disorder, $\alpha<\alpha_{c}\simeq 0.495$, we obtain $\theta\simeq +1.0$ indicating
a superconducting phase for $T>0$. For strong disorder, $\alpha>\alpha_{c}$, the
stiffness exponent $\theta\simeq -1.0$ indicating a non superconducting phase for $T>0$.
\newline\indent
In an applied field type II superconductors have a fixed density of vortex
lines which, with weak disorder, form a distorted Abrikosov lattice or Bragg
glass \cite{Giamarchi:94;Giamarchi:95} at small $T$. Increasing the field
effectively increases the disorder and the Bragg glass phase transforms into
an entangled array of vortex lines with no remnant of translational order. The model
of Eq. (\ref{vortex}) contains the standard description of a vortex lattice as an elastic
periodic solid in a random pinning potential when $\lambda>0$. The essential superconducting
properties seem to be unchanged as $\lambda\rightarrow 0$ but the interaction of the flux lines
determining the periodic lattice structure vanishes when $\lambda =0$. Our strategy is to
impose the required periodic vortex lattice structure by imposing a periodic array of columns of bonds
along $\bf\hat z$ which favor flux lines on these columns. We make the plausible conjecture that
this procedure retains the essential physics of a disordered superconductor at $T\simeq 0$ and it is
the only system which can be simulated with large size $L$. To compute the defect energy
$\Delta E(L)$ it is essential to impose periodic boundary conditions in all directions and imposing
a fixed number of flux lines by the source/target method \cite{Rieger:98} is not applicable.
\begin{figure}
\includegraphics[width=7.0cm,height=7.0cm, angle=-90]{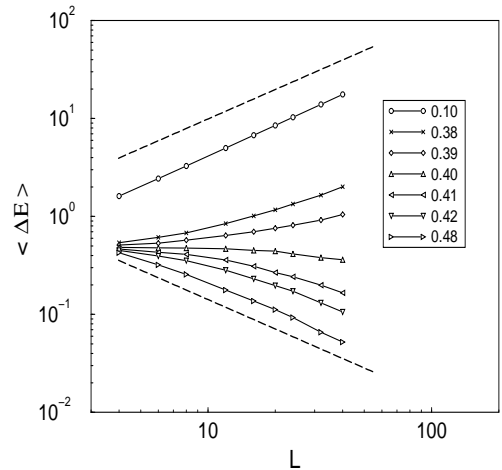}
\caption{$L$ dependence of domain wall energy $\langle\Delta E\rangle$
for finite external field $h=0.25$. Varying the disorder strength
(see legend) changes the sign of the stiffness exponent $\theta$.
Dashed lines have slopes $\pm 1$, while solid lines
are guides for the eye.}
\label{finite_field}
\end{figure}
\newline\indent
We implement the periodic lattice potential by first adding a uniform
$h=Ba_{0}^{2}/\Phi_{0}$ to all $\bf\hat z$ bonds. To ensure that the vortex density
does correspond to this $h$, we introduce a potential $\Delta$ on the bonds of the
expected vortex lattice by $h\rightarrow h+\Delta$ in Eq. (\ref{Hamil_simpl}). The
costs of the flux lines are decreased by increasing $\Delta$ until the matching is
satisfied with one flux line per favorable bond. We thus obtain a ground state of
energy $E_{0}(L)$ with the required number $N=hL^{2}$ of flux lines. The defect
energy $E_{D}(L)$ is obtained in the standard way \cite{Pfeiffer:99,Kisker:98} by
reducing costs until an extra flux loop is induced. This is the simplest way of
introducing the essential periodic part of the random pinning potential on the flux
lines \cite{Giamarchi:94;Giamarchi:95}. It is interesting to note that
our model is essentially that of many elastic lines in a periodic potential
\cite{Knetter:01} so the transition at $\alpha_{c}(h)$ is a transition from a
superconducting state to a state which is both rough and entangled. This is in contrast
to a superconductor in a field with no disorder where the vortex lattice first loses
translational order and becomes a liquid of rigid lines and these lines become
entangled at larger field or temperature \cite{nguyen:99;hove:99}.
\newline\indent
In Fig. (\ref{finite_field}) we show the behavior of the defect energy with system size
$L$ for a fixed value of external field $h = 1/4$ for different disorder strengths $\alpha$.
It is more convenient to fix the field and vary $\alpha$because the period of the vortex
lattice must be commensurate with the system size $L$ which allows only restricted values
of $h=NL^{-2}$ with $N,L$ integers. At small disorder, $\alpha < 0.40$, we observe a
positive stiffness exponent, which  asymptotically tends to $\theta = 1$. Increasing the
disorder, $\alpha > 0.40$, we find that the defect energy decreases with $L$
and for large $L$, $\langle\Delta E(L)\rangle \sim L^{-1}$. There is a critical value
$\alpha_{c}(h=0.25) \simeq 0.40$ separating an ordered from a disordered phase.
Fig. (\ref{ground_state}) shows typical ground state configurations of the system. Below
the critical disorder, Fig. (\ref{ground_state}) left, the lowest energy configuration forms
an almost perfect vortex lattice, while above the critical disorder, Fig. (\ref{ground_state})
right, the lines are rough and entangled \cite{Knetter:01}, as expected for a phase with
proliferation of dislocations. A similar analysis for different values of $h$ shows that
$\alpha_{c}(h)$ decreases monotonically as $h$ increases as shown in Fig. (\ref{alpha.vs.field}).
We estimate the critical value of the external field as $h_c = {\cal O} (1)$ where
$h\equiv Ba_{0}^{2}/\Phi_{0}$ is the mean flux per plaquette. Here $B$ is the actual field,
$a_{0}$ is the lattice spacing and $\Phi_{0} = 2\times 10^{-7}$ Gauss${\bf\cdot}$cm$^2$ is the
flux quantum.
In the strong screening limit, $a_{0}\sim\lambda$, whose
typical value in high-$T_c$ superconductors is $\lambda \simeq 10^{-5}$ cm. This yields an
estimate for the critical field $B_{c}\simeq 10^3$ Gauss. This is to be compared with the
typical experimental value of $B_{c} \simeq 500$ Gauss in BSSCO \cite{Cubitt:93;Klein:01}.
However, this should also hold for YBCO which has a comparable $\lambda$ at $T=0$
\cite{Blatter:94;Natterman:00} and our estimate of $B_{c}$ depends on $\lambda$ only. This
seems to disagree with experimental estimates of $B_{c}\sim 10^{4}$ Gauss \cite{Nishizaki:00}
but the measurements are for $T>50^{o}K$.
\newline\indent
In this Letter we have studied the strongly screened vortex glass in the presence of disorder
and, for the first time, have successfully implemented a defect energy scaling study of the
stability of superconductivity as indicated by the stiffness exponent $\theta$ in an external
field using {\it periodic} boundary conditions. It would be interesting to apply this method
to probe directly the behavior of the model allowing for dislocations in the vortex lattice
by applying appropriate boundary conditions \cite{middle}. It would also be interesting
to treat a more realistic model with finite screening and to show more convincingly
that the $T = \lambda =0$ limit studied in this Letter is the stable renormalization group
fixed point of a disordered superconductor at low temperature.
\newline\indent
We thank H. Rieger and F. O. Pfeiffer for comments and suggestions and P. Mazzanti
for computational assistance. Computations were performed using the Condor
system of Istituto Nazionale Fisica Nucleare. C.G. was supported by ``Fondazione A.
Della Riccia''.

\end{document}